\documentclass[aps,twocolumn,superscriptaddress,showkeys,prb]{revtex4-1}

\usepackage{amsmath}
\usepackage{graphicx}
\usepackage[colorlinks=true,linkcolor=blue,urlcolor=blue,citecolor=blue]{hyperref}

\makeatletter
\renewcommand{\fnum@figure}{Figure \thefigure}
\makeatother

\begin{document}

\title{Direct amplitude-phase near-field observation of higher-order anapole states}

\author{Vladimir A. Zenin}
\affiliation{SDU Nano Optics, University of Southern Denmark, Campusvej 55, DK-5230 Odense M, Denmark}
\email{zenin@mci.sdu.dk}

\author{Andrey B. Evlyukhin}
\affiliation{Laser Zentrum Hannover e.V., 30419 Hannover, Germany}
\email{a.b.evlyukhin@daad-alumni.de}
\affiliation{ITMO University, Kronverksky Pr. 49, St. Petersburg 197101, Russia}

\author{Sergey M. Novikov}
\author{Yuanqing Yang}
\affiliation{SDU Nano Optics, University of Southern Denmark, Campusvej 55, DK-5230 Odense M, Denmark}

\author{Radu Malureanu}
\affiliation{Department of Photonics Engineering, Technical University of Denmark, 2800 Kgs. Lyngby, Denmark}
\affiliation{National Centre for Micro- and Nano-Fabrication, Technical University of Denmark, 2800 Kgs. Lyngby, Denmark}

\author{Andrei V. Lavrinenko}
\affiliation{Department of Photonics Engineering, Technical University of Denmark, 2800 Kgs. Lyngby, Denmark}
\affiliation{ITMO University, Kronverksky Pr. 49, St. Petersburg 197101, Russia}

\author{Boris N. Chichkov}
\affiliation{Laser Zentrum Hannover e.V., 30419 Hannover, Germany}
\affiliation{Leibniz University, 30167 Hannover, Germany}

\author{Sergey I. Bozhevolnyi}
\affiliation{SDU Nano Optics, University of Southern Denmark, Campusvej 55, DK-5230 Odense M, Denmark}

\date{\today}


\begin{abstract}
Anapole states associated with the resonant suppression of electric-dipole scattering exhibit minimized extinction and maximized storage of electromagnetic energy inside a particle. Using numerical simulations, optical extinction spectroscopy and amplitude-phase near-field mapping of silicon dielectric disks, we demonstrate high-order anapole states in the near-infrared wavelength range (900-1700 nm). We develop the procedure for unambiguously identifying anapole states by monitoring the normal component of the electric near-field and experimentally detect the first two anapole states as verified by far-field extinction spectroscopy and confirmed with the numerical simulations. We demonstrate that higher-order anapole states possess stronger energy concentration and narrower resonances, a remarkable feature that is advantageous for their applications in metasurfaces and nanophotonics components, such as non-linear higher-harmonic generators and nanoscale lasers.
[This document is the unedited Author's version of a Submitted Work that was subsequently accepted for publication in \textit{Nano Letters}, \copyright American Chemical Society after peer review. To access the final edited and published work see \url{http://dx.doi.org/10.1021/acs.nanolett.7b04200}.]
\end{abstract}

\maketitle

\begin{figure}
\centering\includegraphics{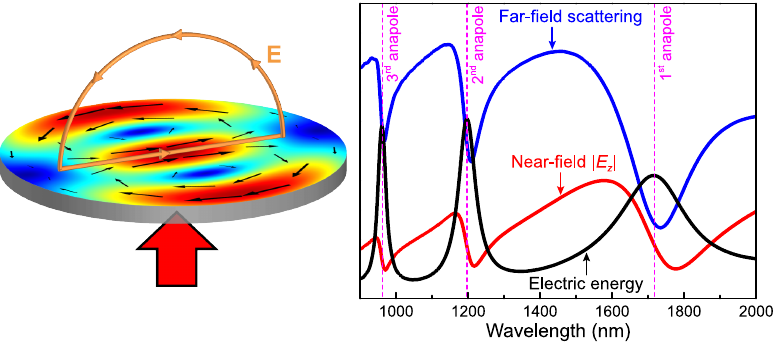}
\end{figure}

\begin{figure*}
\centering\includegraphics{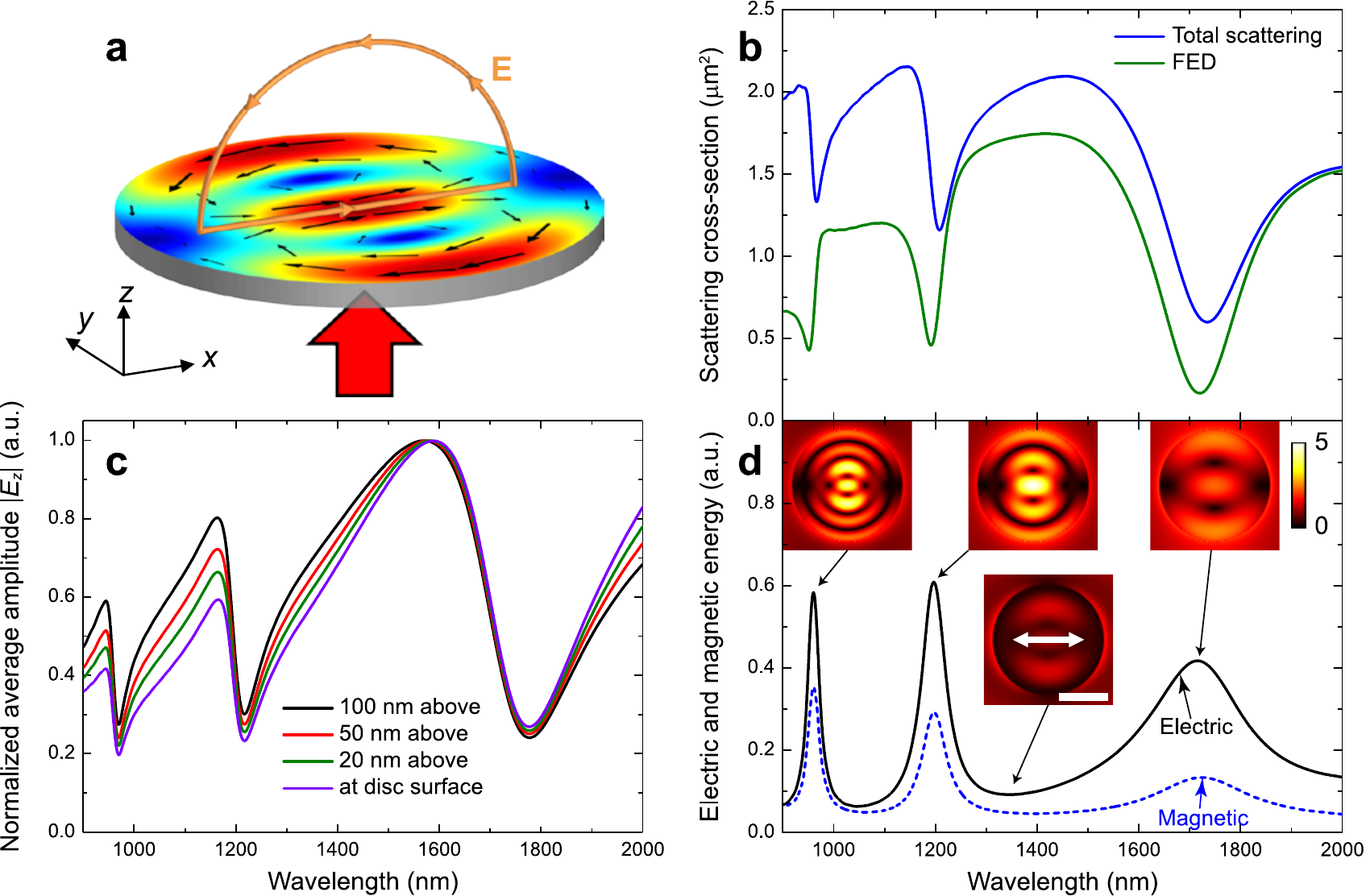}
  \caption{Anapole states. (a) Illustration of the anapole state excitation in a Si disc upon normal $x$-polarized plane-wave illumination. Overlay: $\left|{\bf E}\right|$-field distribution inside the disc at the first anapole state, with arrows representing in-plane electric field direction. At anapole state the electric dipole moment is reduced, causing reduction in both near- and far-field scattering. (b) Full electric dipole (FED, green) and total scattering cross-section (blue) for an 80-nm-thick silicon disc with a diameter of 1100 nm. (c) Normalized amplitude $\left|E_z\right|$, averaged over the disc area and calculated at disc surface (violet) and at the altitude of 20 nm (green), 50 nm (red), and 100 nm above the disc surface (black). (d) Total electric ($\varepsilon_0\varepsilon\iiint\left|{\bf E}\right|^2dV$, solid black) and magnetic energy ($\mu_0\mu\iiint\left|{\bf H}\right|^2dV$, dashed blue), accumulated inside the 1100 nm disc. 1 a.u. corresponds to $2\cdot 10^{-26}$~J at the irradiance of $1~\rm W/m^2$. Insets show $\left|{\bf E}\right|$-field distribution inside the disc, normalized to the amplitude of incident field and calculated at $\lambda$ = 953, 1191, 1340, and 1723 nm (the scale bar of 500 nm). The white double arrow shows the polarization of the incident wave.}
  \label{fig1}
\end{figure*}

Recently, high-refractive index dielectric particles attracted considerable attention of nanophotonics community due to possibility to support electric and magnetic optical resonances \cite{diel00, Vesperinas1, diel0, diel1, diel2, diel3, Vesperinas2, diel4, diel5, diel6}. In order to understand their radiation scattering behavior, one can either decompose scattering electromagnetic fields outside the particle (known as decomposition in spherical multipoles\cite{spher_mult}), or decompose the electric field inside a particle (known as decomposition in Cartesian multipoles\cite{Cartesian_mult}). The above two approaches have many things in common; they even share similar notions such as electric dipole, which might cause a confusion. Moreover, spherical multipole decomposition can be done using only fields inside the particle and the near-to-far-field relation \cite{Kaivola, Rockstuhl}. The spherical multipole family contains ordinary electric/magnetic dipole, quadrupole, octupole, etc., like in the Mie theory \cite{Mie}, while the Cartesian multipole family includes other additional terms, which are determined by special field distributions inside the particle but have the same far-field scattering diagram as the above mentioned spherical multipoles. For example, a toroidal dipole moment is the third-order Cartesian multipole moment radiating electromagnetic waves in all wave zones as the Cartesian electric dipole \cite{toroid, ChanNP_2011, EvlyukhinPRB2016}. To avoid any confusion, we will call spherical multipoles as full multipoles (for example, full electric dipole), because every spherical multipole, generating a certain scattering diagram, contains contributions of all Cartesian multipoles with the same scattering pattern. Explicit connections between several first spherical multipoles (dipole and quadrupole terms) and corresponding Cartesian multipoles for arbitrary shaped particles were recently considered in Ref.~\citenum{Rockstuhl}. It was shown that a toroidal dipole moment can give comparable contribution to the scattering as the electrical dipole moment, therefore it can either enhance \cite{EvlyukhinOL2017} or suppress \cite{Zheludev, anapole} the full electric dipole (FED) scattering. In certain conditions it can cancel completely the FED scattering, which was observed for dielectric spheres and called the anapole mode \cite{anapole}. A similar case happens in elementary particle physics, where the anapole term was first introduced as a non-radiating source \cite{elem}.  However, the ideal non-radiating anapole cannot be excited by a propagating optical wave due to the reciprocity theorem. To our knowledge, a pure anapole was excited only by a standing wave with a specific field profile \cite{anapole2}. However, for many practical applications it is more useful to study conditions when the scattering by a particle is substantially suppressed under plane-wave illumination. For most shapes of small dielectric particles the scattering is dominated by the electric dipole moment, therefore we will call this anapole state, which is defined as a condition when the FED scattering is at the local minimum. In general case the anapole state is caused by the destructive interference between electric, toroidal, and other electric-dipole-type moments. It was shown that for high refractive index discs with high diameter-to-thickness aspect ratio at the anapole state there is a drop in the total scattering cross-section and, at the same time, substantial boost of the electromagnetic energy inside the particle \cite{Ge_discs1, Ge_discs2, NC17}. Since the electric dipole radiation is suppressed in the anapole state, it is reasonable to assume that there should be a signature in the near-field distribution, corresponding to the far-field suppression \cite{spher_mult}. Here we demonstrate, for the first time to our knowledge, such near-field detection of the anapole state. It is revealed that the normal component of the electric near-field drops down at such states. Note that the near-field detection shouldn't be confused with the near-field mapping of the anapole state (like the one demonstrated recently in Ref.~\citenum{anapole}). The difference is the following: the near-field mapping itself shows only correlation with the simulated near-field distribution and does not allow direct identification of anapole states, whereas the near-field detection method presented in this work provides a straight-forward tool for identification and observation of anapole states. This method is valid for different illumination wavelengths and different orders of anapole states.

The paper is composed as following. First, we conduct simulations of isolated Si discs in free space and show the existence of different-order anapole states. They are defined by the drop in the FED scattering and are accompanied by a peak in the electromagnetic energy concentration inside the particle and a drop in both the total scattering cross-section and the amplitude of the normal electric near-field component, measured at a certain height above the disc and averaged over the disc area.  We also demonstrate that higher-order anapole states feature higher energy accumulation and narrower resonances, compared to the first (fundamental) anapole state. We make a far-field experimental detection of anapole states by measuring the transmission and correspondent extinction spectra for a variety of disc diameters. Then, we proceed to investigation of the Si discs in the amplitude- and phase-resolved scanning near-field optical microscope (SNOM). Using a simple symmetry-based procedure, we extract the normal component of the near-field $E_z$. The drop in $\left|E_z\right|$, averaged over the disc area, is then used for the experimental near-field identification of anapole states. Finally, we summarize our findings and show that the anapole states can be detected by the drop in near-field $\left|E_z\right|$ and/or in the far-field extinction, or by the peak in the electric energy accumulated inside the particle.

\section*{Results and discussion}

We begin with numerical simulations of an isolated 80-nm-thick Si disc with 1100 nm diameter in free space, illuminated normally with an $x$-polarized plane wave (Figure~\ref{fig1}a). First we compute the total scattering of the disc for the near-infrared wavelength range of 900-2000 nm (see Methods), which features three dips (Figure~\ref{fig1}b). In order to understand scattering spectra, we apply the spherical multipole decomposition (by using calculated E-field inside the disc and \cite{Rockstuhl}) and find the accompanying dips in the contribution of the full electric dipole (FED) scattering, which we use to define the anapole states. The decomposition into Cartesian multipoles verifies that the first anapole state (at $\lambda \approx 1700$~nm) is indeed caused by destructive interference of the toroidal and electric dipole moments (Supporting Information, Figure~S1). However, the Cartesian multipole decomposition fails to describe accurately higher-order anapole states, appearing at shorter wavelength (at $\lambda \approx 1200$ and 950 nm, correspondingly), since it intrinsically assumes a small parameter of $2\pi r/\lambda$. Thus, higher-order anapole states should be described with the spherical multipole decomposition, and the dip in FED scattering is not simply a result of destructive interference between electric and toroidal dipole moments. The similarity between total and FED scattering shows that the far-field scattering is dominated by the electric dipole moment (Figure~\ref{fig1}b). Thus, it is natural to assume that the near-field is also dominated by the same electric dipole. We found that it is indeed so for the normal $E_z$ component of the near field. When its amplitude is calculated at a certain altitude above the disc and averaged over the disc area, than it also features three dips, corresponding to anapole states (Figure~\ref{fig1}c). This near-field method of anapole identification is robust to the altitude, where $E_z$ is calculated. Later we use the altitude of 50 nm, since it equals to the oscillation amplitude of a SNOM probe in the experiment. Finally, we compute the electric and magnetic energy inside the disc ($\varepsilon_0\varepsilon\iiint  \left|{\bf E}\right|^2dV$ and $\mu_0\mu\iiint  \left|{\bf H}\right|^2dV$), which now features three pronounced peaks at the anapole states (Figure~\ref{fig1}d). This correlation can be intuitively understood by the increase of the quality factor of the resonator, causing higher energy accumulation inside the resonator resulted from suppressed scattering. It should be noted that higher-order anapole states possess stronger energy concentration and narrower resonances, both indicating their higher quality factor compared to the first (fundamental) anapole state. The insets in Figure~\ref{fig1}d show the amplitude of the electric field inside the disc at different illumination wavelengths. At the first anapole state (also shown in Figure~\ref{fig1}a) the electric field distribution features three hot spots with alternating direction of the in-plane electric field, forming two opposite vortices, which generate a magnetic quadrupole \cite{anapole}. Thus the outer part of the disc has the opposite direction of the in-plane electric field relative to the central part, which results in a reduced electric dipole moment, while electric field amplitude is high inside the disc. At higher-order anapole states the number of {\bf E}-field hot spots increases \cite{Ge_discs2}, similarly to the standing-wave patterns formed in brick antennas at high-order modes \cite{Yuanqing}.

\begin{figure}[b]
\centering\includegraphics{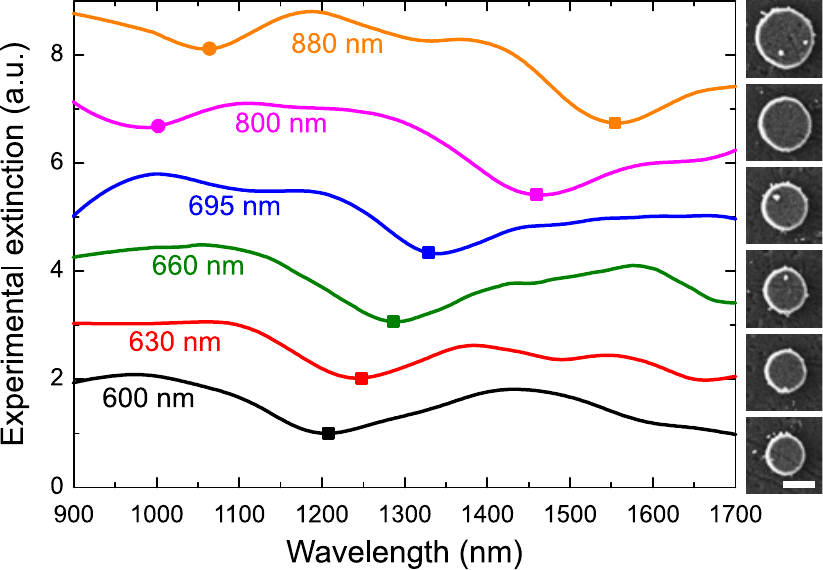}
  \caption{Experimental far-field detection of anapole states. Experimental extinction spectra of isolated silicon disks with a diameter ranging from 600 to 880 nm. The baseline for each spectrum has an offset of 1 a.u. Points show the position of the dip in simulated scattering spectra (squares and circles correspond to the first and second anapole state, respectively). Right: SEM images of corresponding disks (the scale bar of 500 nm).}
  \label{fig2}
\end{figure}

\begin{figure*}[htb]
\centering\includegraphics{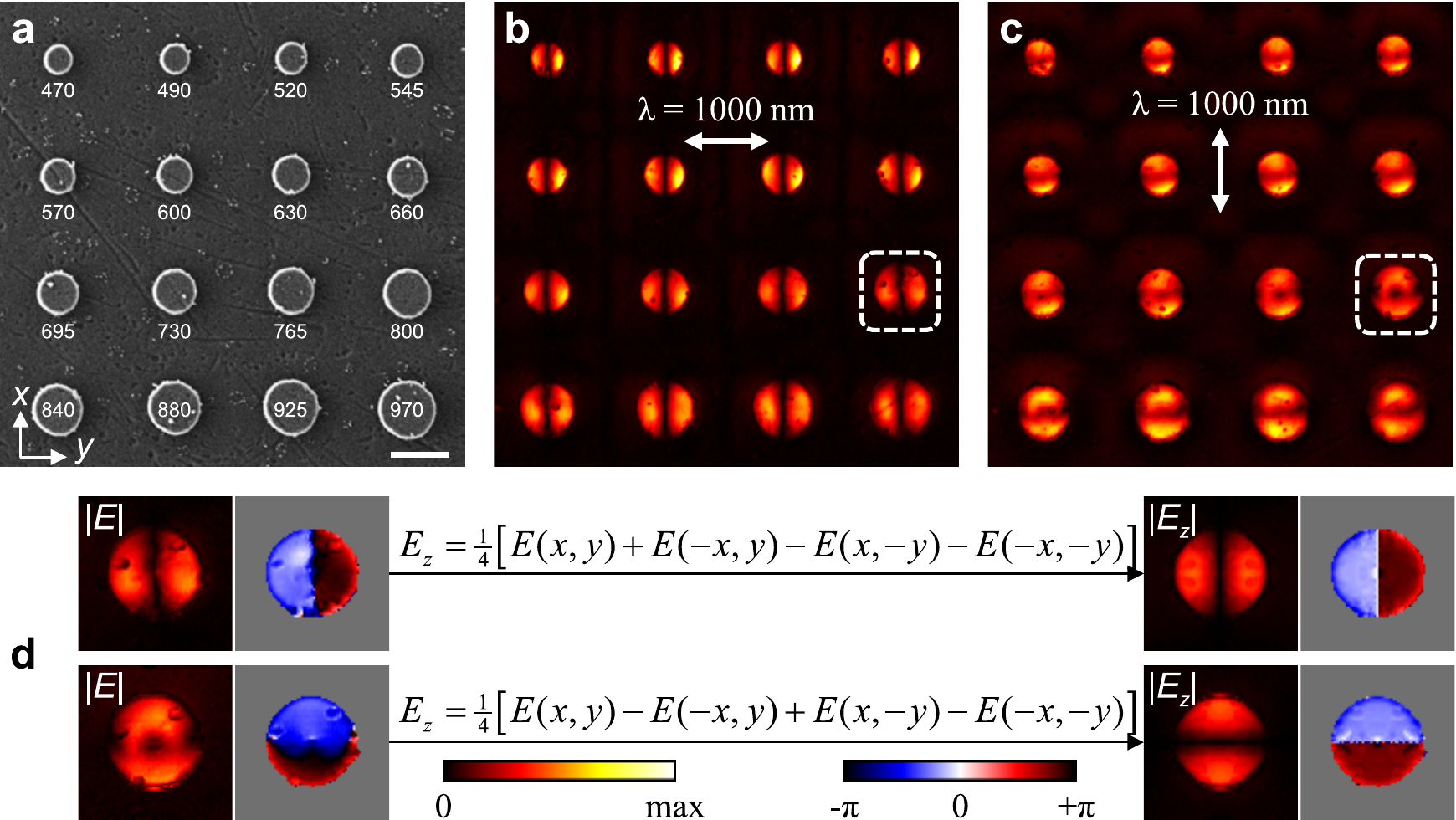}
  \caption{Near-field characterization. (a) SEM and (b, c) pseudocolor SNOM maps of near-field amplitude for silicon disks under normal and linearly polarized illumination (depicted with double arrow). Disk diameters are shown in a. Scale bar, 1 $\mu$m. The disk with the lowest near-field amplitude is encircled with the white dashed line in (b), (c). (d) Processing of near-field maps for 800 nm disk by using symmetry in order to get a clean near-field component $E_z$.}
  \label{fig3}
\end{figure*}

To confirm the far-field response discussed above, we performed a series of experiments to observe and detect the excitation of anapole states. Si discs of 80 nm thickness and diameters varied from 470 to 970 nm were fabricated on a fused silica substrate (see Methods). These discs, separated by 50 $\mu$m to reproduce the isolated condition, were used for far-field transmission measurements (see Methods). The measured transmittance $T$ was used to plot extinction $1-T$, which should be proportional to the total scattering, since absorption is negligible for wavelength above 900 nm (see Supporting Information, Figure~\ref{S2}). Points in Figure~\ref{fig2} show the position of the dip in simulated scattering spectra (squares and circles correspond to the first and second anapole state, respectively), which demonstrates a good agreement between experimental and numerical results.

Then another set of discs with the same diameters, but arranged in a four-by-four array with a period of 2 $\mu$m was investigated with the scattering-type amplitude- and phase-resolved scanning near-field microscope (SNOM) at wavelengths of 900-1640 nm (see Methods). It should be noted that SNOM transfer function is rather complicated, on the one hand because SNOM probe is not significantly small compared to a studied object, and on the other hand because of the sophisticated detection process, when the scattered signal is first modulated due to the tip's oscillations and then demodulated to suppress constant background. However, in previous studies it was a good enough agreement between SNOM maps and simulated normal component of the near-field $E_z$, recorded at the altitude of 50 nm above the object without the tip \cite{pizza, slot, strip}. The in-plane near-field components $E_x$ and $E_y$ are taken into account only in special cases \cite{Boosting}. It is already clear from the SNOM amplitude distribution (Figure~\ref{fig3}b,c) that the certain disc size corresponds to the minimum average near-field amplitude (it is encircled with the white dashed line and its diameter was 800 nm for the incident wavelength of  $\lambda$ = 1000 nm). However, there is a clear difference between SNOM maps for the $x$- and $y$-polarized incident beams. This discrepancy can be explained by anisotropy of the tip, which results in different sensitivity to $E_x$ and $E_y$ near-field components. However, assuming that the influence of the components is linear summed up in the total influence, and taking into account different symmetry of each near-field component along $x$- and $y$-axes, it is possible to decompose the recorded near-field $E$ into $E_x$, $E_y$, and $E_z$ (see Methods). This processing method, applied to different SNOM maps for both polarizations of the incident beam, resulted in nearly the same $E_z$ distribution (Figure~\ref{fig3}d).

\begin{figure}
\centering\includegraphics{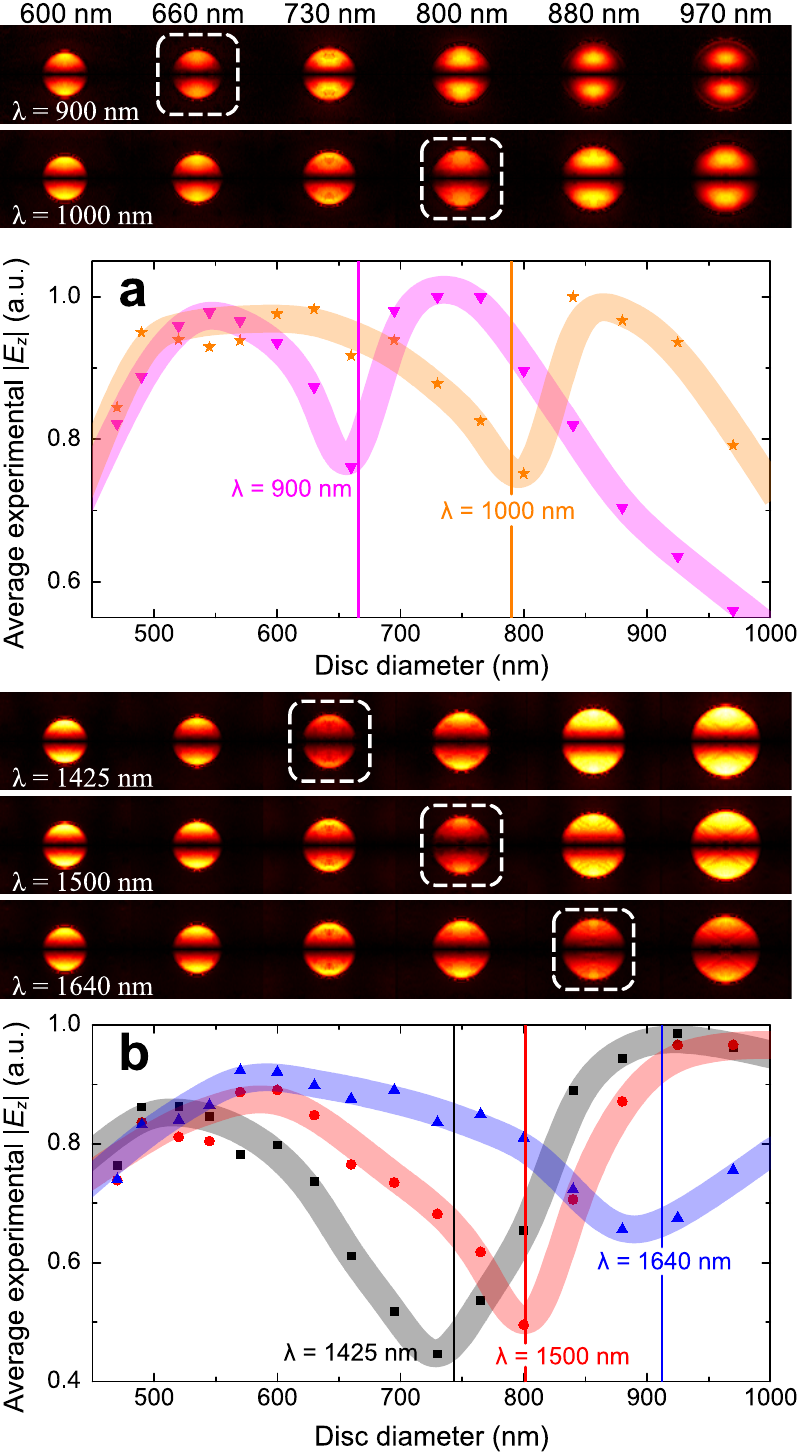}
  \caption{Near-field detection of anapole states. Experimental near-field $\left|E_z\right|$ distribution and its average value for silicon disks, measured at different illumination wavelength: (a) 900 (magenta down triangles), 1000 (orange stars), (b) 1425 (black squares), 1500 (red circles), and 1640 nm (blue up triangles). The thick transparent curves are for eye guidance. Disk sizes are labeled on top. The disk with the lowest near-field amplitude is encircled with the white dashed line. Vertical lines in graphs show the disk diameter with the lowest simulated $\left|E_z\right|$ and are related to the (b) first-order and (a) second-order anapole states.}
  \label{fig4}
\end{figure}

Then we performed such decomposition for near-field maps of all disc sizes, recorded at five different illumination wavelengths: 900, 1000, 1425, 1500, and 1640 nm (Figure~\ref{fig4} and Supporting Information, Figures \ref{S5}-\ref{S6}). It was found that there is a clear drop in $\left|E_z\right|$ for a certain disc size, which in turn depends on the wavelength. Simulations confirm the drop in the average near-field $\left|E_z\right|$, calculated 50 nm above the disc surface (the position of the dip is shown with vertical lines in Figure~\ref{fig4}a,b). According to simulations, such drop appears both for first and second anapole states. 

\begin{figure}
\centering\includegraphics{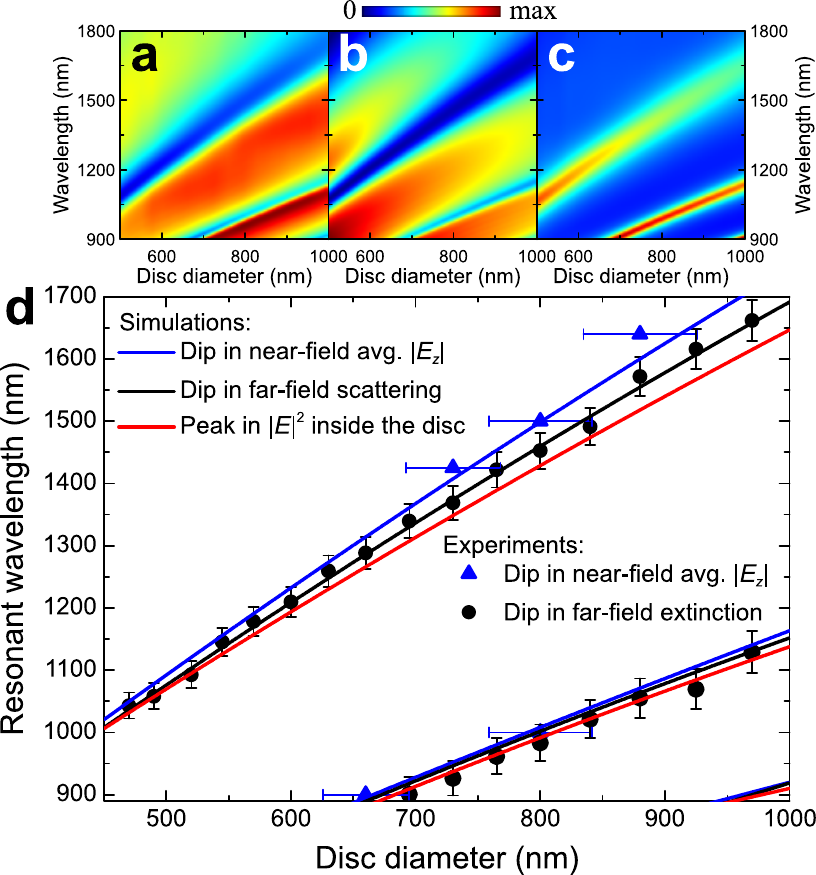}
  \caption{Various observation methods of anapole states. (a-c) Simulated (a) average near-field $\left|E_z\right|$, (b) total scattering cross-section, normalized to the disc cross-section area, and (c) average electric energy density inside the disc for silicon discs with a diameter ranging from 500 to 1000 nm. (d) Resonance position, experimentally measured and calculated with different techniques. Error bars are evaluated from measurements of different samples (for far-field extinction) and from the discrete set of disc diameters (for near-field).}
  \label{fig5}
\end{figure}

To sum it up, we compare different techniques, used for the identification of anapole states. First we plot simulated average near-field $\left|E_z\right|$ (Figure~\ref{fig5}a), calculated 50 nm above the top disc surface, and the total scattering cross-section, normalized to the disc cross-section area (Figure~\ref{fig5}b), for different wavelengths and disc diameters. Even though they represent either near- of far-field properties, they both look very similar, with dips corresponding to the anapole states. At such states there is a clearly pronounced peak in the total electric energy inside the disc, normalized to its volume (Figure~\ref{fig5}c). Finally, Figure~\ref{fig5}d shows positions of the extrema, found with the above mentioned techniques, and the experimental findings - dips in extinction spectra for each disc size and dips in average near-field $\left|E_z\right|$ for each selected wavelength. As can be seen, there is a good agreement between experimentally measured and simulated results. Also it was found that different techniques provide almost the same position of the anapole states with deviation within 5\%. The origin of such deviation is the well-known mismatch between the peak positions in the near- and far-fields \cite{Near_far_peak}.

\section*{Outline}

In summary, we have thoroughly investigated the anapole states of Si discs in free space and on top of a glass substrate. We defined an anapole state as a condition when the full electric dipole contribution to the scattering cross-section is resonantly suppressed. This suppression originates from destructive interference of electric-dipole-like moments corresponding to the Cartesian multipole decomposition, i.e., electric and toroidal dipoles for the first-order anapole state. We have found that the FED scattering at higher-order anapole states, appearing for larger disc sizes or at shorter wavelengths, cannot be described as simply destructive interference of only electric and toroidal dipoles, since the high-order Cartesian electric-dipole-like moments cannot be neglected. We have shown that the anapole states in the discs can be identified by a drop in the total scattering cross-section and a local maximum in the total accumulated electromagnetic energy inside the particles. Additionally we have found that the anapole states also feature a reduction of the normal near-field electric component $\left|E_z\right|$. This property has been successfully used for experimental identification of the anapole states. All three identification techniques are found to provide similar results with deviations less than 5\%. The near-field detection of anapole states can be particularly useful for sets of particles with subwavelength variations, for example in metasurfaces, when other far-field techniques might fail due to the far-field resolution limitations. Finally, we have demonstrated that higher-order anapole states possess stronger energy concentration and narrower resonances, a remarkable feature that is advantageous for their applications in nanophotonics components, such as non-linear higher-harmonic generators and nanoscale lasers.

\section*{Methods}

{\bf Fabrication.} The disks were fabricated in silicon deposited on silica wafers. The new silica wafers were first cleaned using a standard RCA clean, without the HF steps. Thereafter, 80 nm of amorphous Si was LPCVD grown. The thickness and refractive index of the grown Si was measured using a FilmTek 4000TM spectrometer. The structures were then defined using e-beam lithography. Since the standard resist used is a positive one, we used a combination of lift-off and etching to define them. Thus, after the spinning of the AR-P 6200.09 resist and the definition of the structures, a thin 50 nm layer of Al was deposited on the sample. Using a standard lift-off technique, we removed the rest of the resist and the Al layer deposited on top of it. This way, we obtained an Al hard mask for etching the Si. We used a BOSCH process to etch the 80 nm of silicon and define the structures. The last step involved removing of the Al such that the structures are now defined only in Si. A quick overview of the fabrication steps can be seen in Supporting Information, Figure~\ref{S2}. Designed disc diameters were 30 nm larger than those reported in this work (i.e., from 500 to 1000 nm). Then upon comparing the experimental and simulated anapole positions (Figure~\ref{fig5}d), we found that the best agreement will be if we assume a 30 nm reduction in the fabricated disc diameter. This assumption is appropriate, since it is only a 3-6\% reduction, and it is comparable with the Si thickness (about 3 times smaller). Also it should be noted that the size of the fabricated discs cannot be measured more accurately neither with AFM nor with SEM (the latter is because of charging problems of non-conducting samples). Only this assumption was used as a fitting parameter, all other parameters (Si thickness and its dielectric permittivity) were taken from experimental measurements.

{\bf Numerical simulations.} Scattering spectra and field distributions were calculated using a three-dimensional finite-difference time-domain (FDTD) method with a commercial software package (Lumerical). A simulation box of $3\times3\times3~\rm \mu m^3$ was used with the perfectly matched layer conditions on every boundary and a mesh size of 5 nm over the volume of the silicon disk. The silicon disk was excited by a normal-incident total-field/scattered-field plane wave source ranging from 900 to 2000 nm. The scattered power was then directly calculated by means of the transmission monitors surrounding the disk and the substrate. Multipole decomposition was carried out in both Cartesian \cite{mult1, Yuanqing} and spherical \cite{Rockstuhl} basis. All calculations used the measured permittivity value of Si (Supporting Information, Figure~\ref{S2}), the refractive index of glass was set to 1.45.

{\bf Far-field spectroscopy.} Optical properties of the fabricated Si discs were studied using spatially resolved linear transmission spectroscopy. The spectroscopic transmission analysis was performed on a BX51 microscope (Olympus) equipped with a halogen light source and fiber-coupled grating NIR spectrometer NIRQuest (Ocean Optics) with the wavelength resolution of 7.2 nm. The light was collected in the transmission configuration using the MPlanFL (Olympus) objective with magnification $\times 100$ (NA = 0.9). By using a pinhole in the image plane, we collected spectra from an area with a diameter of $\sim$20 $\mu$m. The extinction experimental data in Figure~\ref{fig2} represent $1-T_{\rm str}/T_{\rm ref}$, where $T_{\rm str}$ is the transmission spectrum measured from a single disc and $T_{\rm ref}$ is the transmission spectrum recorded from the clean glass surface.

{\bf Near-field microscopy.} We used AFM-based SNOM from NeaSpec with standard platinum-coated Si tips (Arrow NCPt from NanoWorld). The sample was illuminated normally from below, while the scattered signal was detected (Supporting Information, Figure~\ref{S3}). The AFM tip was tapping with frequency $\Omega \sim$ 250 kHz and the amplitude $\sim$50 nm. Our SNOM uses an interferometric pseudoheterodyne detection \cite{pseudohet}, with a reference beam passing through the path with optical length oscillations at $f \sim$ 300 Hz. We detected a demodulated signal of sidebands around the $4^{\rm th}$ harmonic ($4\Omega + nf$, $n$ = 1,2), which was used to produce a complex-valued near-field. The sampling interval was equal to the pixel time = 20 ms. The illumination spot size at the sample surface was estimated to be $\sim$3 $\mu$m.

{\bf SNOM decomposition method.} From simulations it is found that different near-field components of the illuminated disc have different symmetry (see Supporting Information, Figure~\ref{S4}). For example, at $x$-polarization of the incident beam, the near-field component $E_x$ is symmetrical for both axis, i.e., $E_x(-x) = E_x(x)$ and $E_x(-y) = E_x(y)$, while $E_y$ is asymmetrical: $E_y(-x) = -E_y(x)$ and $E_y(-y) = -E_y(y)$. Finally, $E_z(-x) = -E_z(x)$ and $E_z(-y) = E_z(y)$. Therefore, if we assume linear superposition in the recorded field
\[ E(x,y)=C_xE_x(x,y)+C_yE_y(x,y)+ C_zE_z(x,y)\]
then the following operation will give
\begin{equation*}
%\resizebox{.9 \textwidth}{!}
%{$E(x,y)+a_xE(-x,y)+a_yE(x,-y)+a_xa_yE(-x,-y) = (1+a_x)(1+a_y)C_xE_x + (1-a_x)(1-a_y)C_yE_y + (1-a_x)(1+a_y)C_zE_z$}
\begin{array}{r}
E(x,y)+a_xE(-x,y)+a_yE(x,-y) + a_xa_yE(-x,-y)\\
= (1+a_x)(1+a_y)C_xE_x  + (1-a_x)(1-a_y)C_yE_y\\
									 		 + (1-a_x)(1+a_y)C_zE_z
\end{array}
\end{equation*}
Thus, by choosing $a_x = a_y = 1$, the operation will result in the $E_x$ component, $a_x = a_y = -1$ will give $E_y$, and $a_x = -a_y = -1$ will give $E_z$ (see Supporting Information, Figure~\ref{S4}). After such decomposition, applied for both incident polarizations, it was found that our SNOM is sensitive mostly to the $z$ component, moderately to the $x$ component, and almost negligibly to the $y$ component of the near-field (see Supporting Information, Figure~\ref{S4}). Finally, a choice of $a_x = -a_y = 1$ should give zero in the above model, therefore, when applied to recorded SNOM maps, its result can be claimed as an estimate of the error for this decomposition method.

\textbf{Acknowledgments}

The authors acknowledge financial support from the European Research Council, Grant No. 341054 (PLAQNAP); from the University of Southern Denmark (SDU2020 funding); from the Deutsche Forschungsgemeinschaft (Germany), the project EV 220/2-1; and from the Villum Fonden (DarkSILD project). The development of the theoretical and numerical models has been partially supported by the Russian Science Foundation (Russian Federation), the project 16-12-10287.

\bibliography{zenin}

%%%below is supplementary
%\pagebreak
\pagebreak
\widetext
\begin{center}
\textbf{\large Supporting information}
\end{center}
%%%%%%%%%% Merge with supplemental materials %%%%%%%%%%
%%%%%%%%%% Prefix a "S" to all equations, figures, tables and reset the counter %%%%%%%%%%
\setcounter{equation}{0}
\setcounter{figure}{0}
\setcounter{table}{0}
\setcounter{page}{1}
\makeatletter
\renewcommand{\thepage}{s\arabic{page}}
\renewcommand{\theequation}{S\arabic{equation}}
\renewcommand{\thefigure}{S\arabic{figure}}
\renewcommand{\bibnumfmt}[1]{[S#1]}
\renewcommand{\citenumfont}[1]{S#1}
%%%%%%%%%% Prefix a "S" to all equations, figures, tables and reset the counter %%%%%%%%%%
\begin{figure}[hb]
\centering\includegraphics{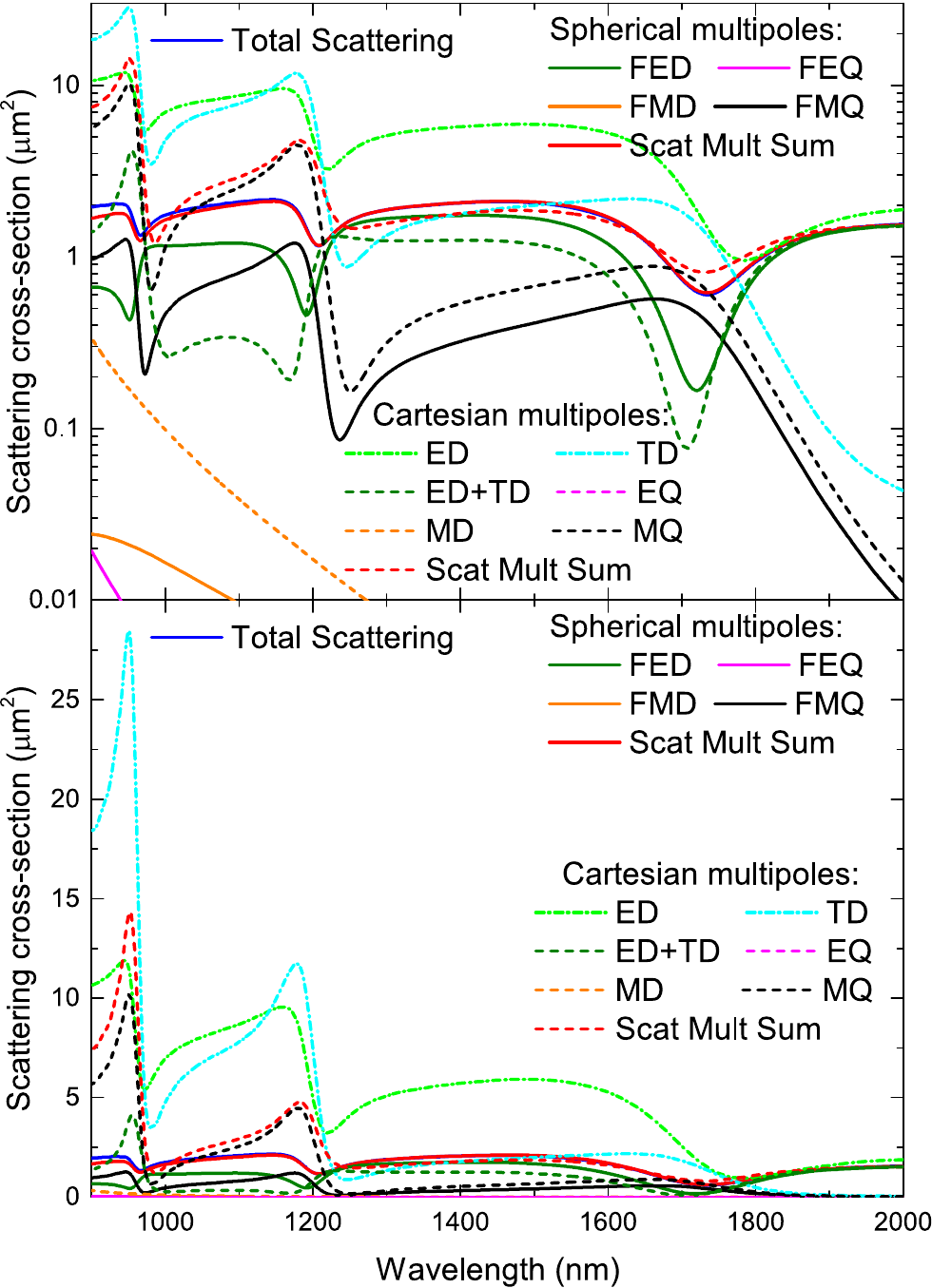}
  \caption{Decomposition of the scattering in Spherical and Cartesian multipoles. As in Figure~\ref{fig1}, we consider scattering by the 80-nm-thick silicon disc with a diameter of 1100 nm in free space. F stand for full, E, M, and T stand for electric, magnetic, and toroidal, D and Q stand for dipole and quadrupole, respectively. Scat Mult Sum is a sum of the scattering contributions of considered multipoles. Given set of Cartesian multipoles describes scattering fairly good only for the first anapole state, while for higher orders (lower wavelength) it is clear that one should take into consideration other higher-order multipoles.}
	\label{S1}
\end{figure}

\begin{figure}
\centering\includegraphics{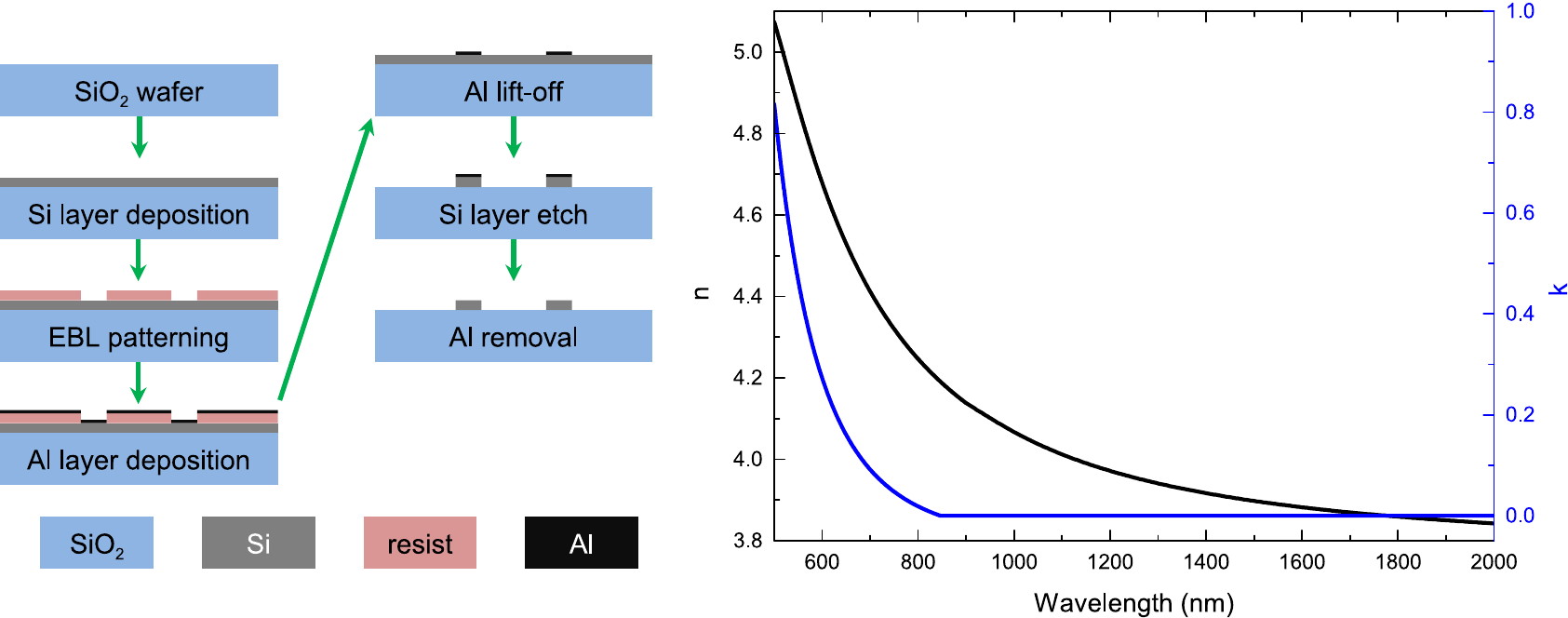}
  \caption{Fabrication process and measured refractive index with extinction coefficient of fabricated 80-nm-thick Si film. The disks were fabricated in silicon deposited on silica wafers. The new silica wafers were first cleaned using a standard RCA clean, without the HF steps. Thereafter, 80 nm of amorphous Si was LPCVD grown. The thickness and refractive index of the grown Si was measured using a FilmTek 4000TM spectrometer. The structures were then defined using e-beam lithography. Since the standard resist used is a positive one, we used a combination of lift-off and etching to define them. Thus, after the spinning of the AR-P 6200.09 resist and the definition of the structures, a thin 50 nm layer of Al was deposited on the sample. Using a standard lift-off technique, we removed the rest of the resist and the Al layer deposited on top of it. This way, we obtained an Al hard mask for etching the Si. We used a BOSCH process to etch the 80 nm of silicon and define the structures. The last step involved removing of the Al such that the structures are now defined only in Si. Designed disc diameters were 30 nm larger than those reported in this work (i.e., from 500 to 1000 nm). Then upon comparing the experimental and simulated anapole positions (Figure~\ref{fig5}d), we found that the best agreement will be if we assume a 30 nm reduction in the fabricated disc diameter. This assumption is appropriate, since it is only a 3-6\% reduction, and it is comparable with the Si thickness (about 3 times smaller). Also it should be noted that the size of the fabricated discs cannot be measured more accurately neither with AFM nor with SEM (the latter is because of charging problems of non-conducting samples). Only this assumption was used as a fitting parameter, all other parameters (Si thickness and its dielectric permittivity) were taken from experimental measurements.}
  \label{S2}
\end{figure}

\begin{figure}
\centering\includegraphics{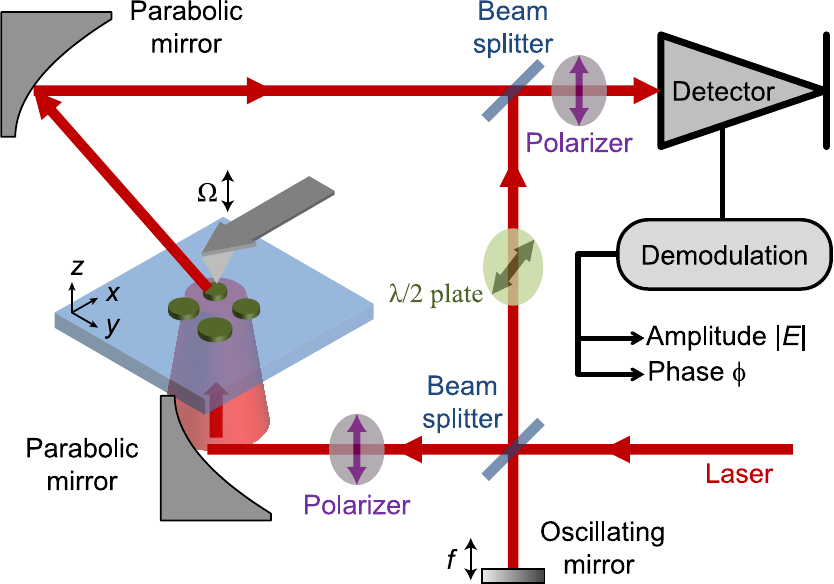}
  \caption{SNOM setup. We used AFM-based SNOM from NeaSpec with standard platinum-coated Si tips (Arrow NCPt from NanoWorld). The sample was illuminated normally from below, while the scattered signal was detected. The AFM tip was tapping with frequency $\Omega \sim$ 250 kHz and the amplitude $\sim$50 nm. Our SNOM uses an interferometric pseudoheterodyne detection, with a reference beam passing through the path with optical length oscillations at $f \sim$ 300 Hz. We detected a demodulated signal of sidebands around the $4^{\rm th}$ harmonic ($4\Omega + nf$, $n$ = 1,2), which was used to produce a complex-valued near-field. The sampling interval was equal to the pixel time = 20 ms. The illumination spot size at the sample surface was estimated to be $\sim$3 $\mu$m.}
	\label{S3}
\end{figure}

\begin{figure}
\centering\includegraphics{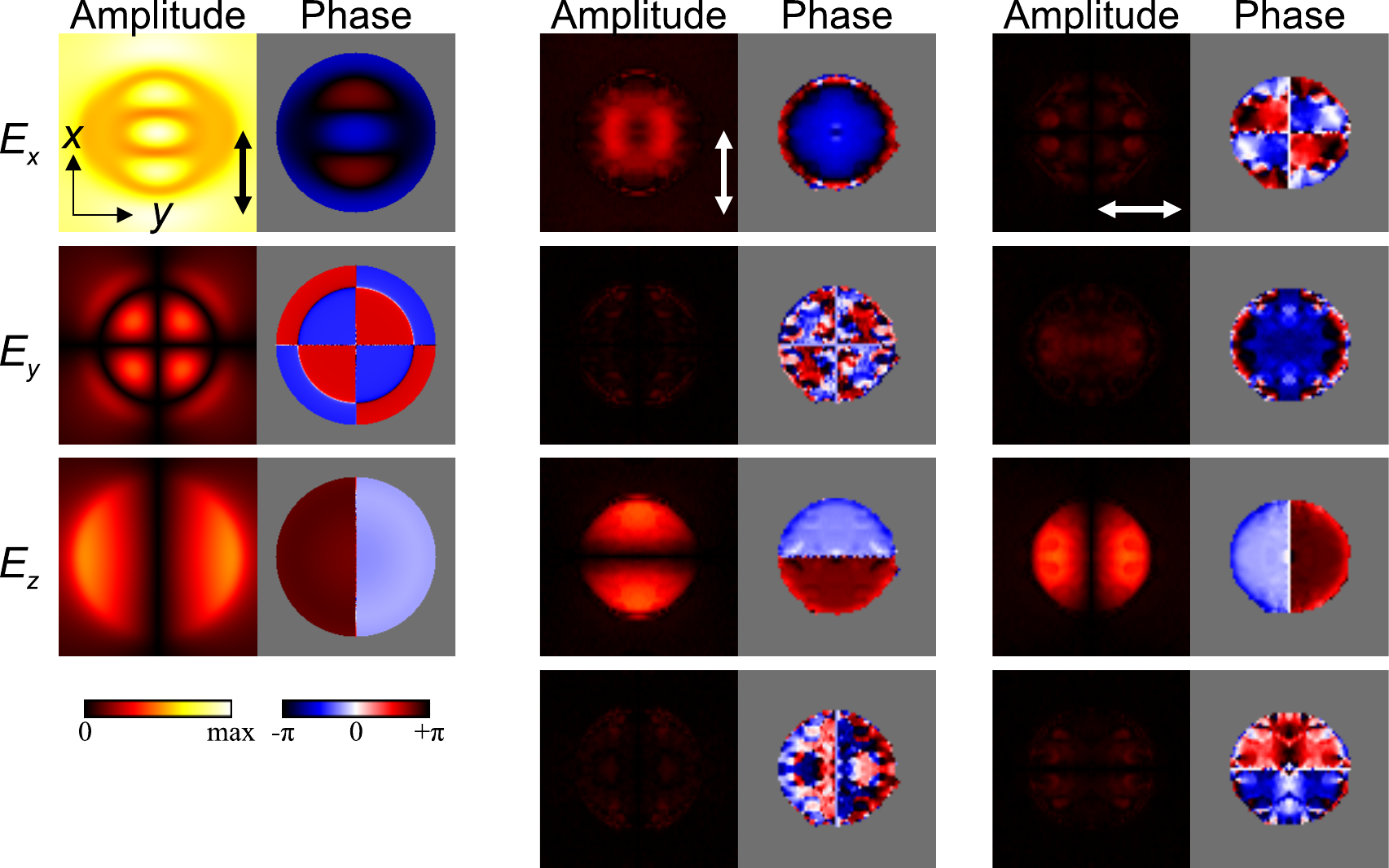}
  \caption{Decomposition into near-field components. Amplitude and phase of near-field components for 80-nm-thick Si disc of diameter 800 nm on a glass substrate, illuminated normally at $\lambda$ = 1000 nm. Left column shows simulations, and the rest two are results of decomposition of recorded SNOM maps shown in Figure~\ref{fig3} of the main text. Double arrow depicts the polarization of the incident beam. First three rows correspond to the $x$, $y$, and $z$ component of the near-field, while the last fourth row of experimental data can be used as an estimate of the error of decomposition method based on linearity of SNOM transfer function. It is clear that $E_x$ component is dominating in simulations, while SNOM is mostly sensitive to the $z$ component, moderately to the $x$ component, and almost negligibly to the $y$ component of the near-field.}
	\label{S4}
\end{figure}

\begin{figure}
\centering\includegraphics{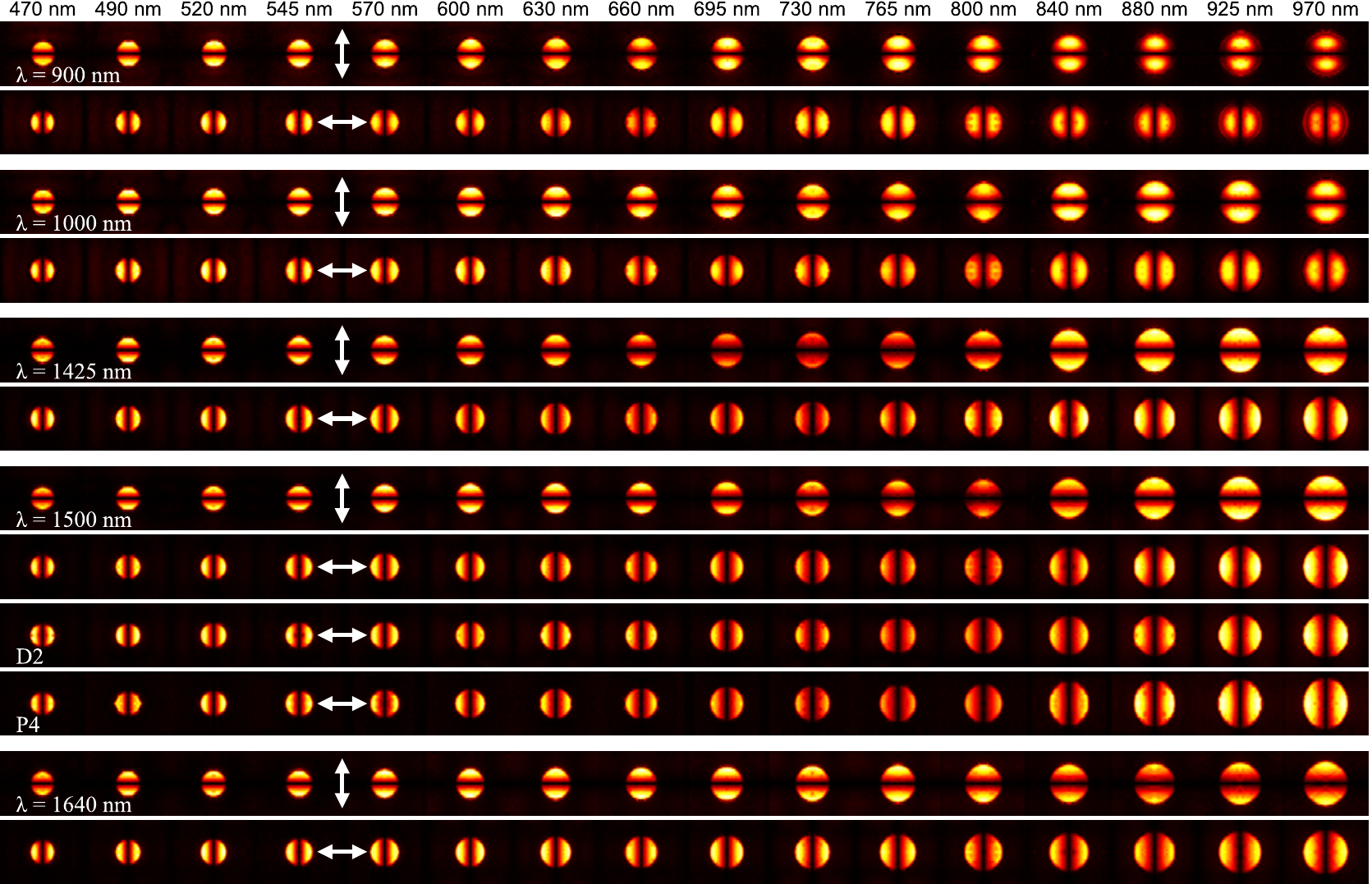}
  \caption{Decomposed measured near-field component $\left|E_z\right|$ for various disc sizes, wavelengths, and polarizations. The polarization of the incident light is illustrated with double arrow. Disc diameter and wavelength are labeled. Two lower near-field maps at $\lambda$ = 1500 nm were done on a duplicated structure (labeled D2) and on a structure where discs were arranged with an increased periodicity of P = 4 $\mu$m (labeled P4).}
	\label{S5}
\end{figure}

\begin{figure}
\centering\includegraphics{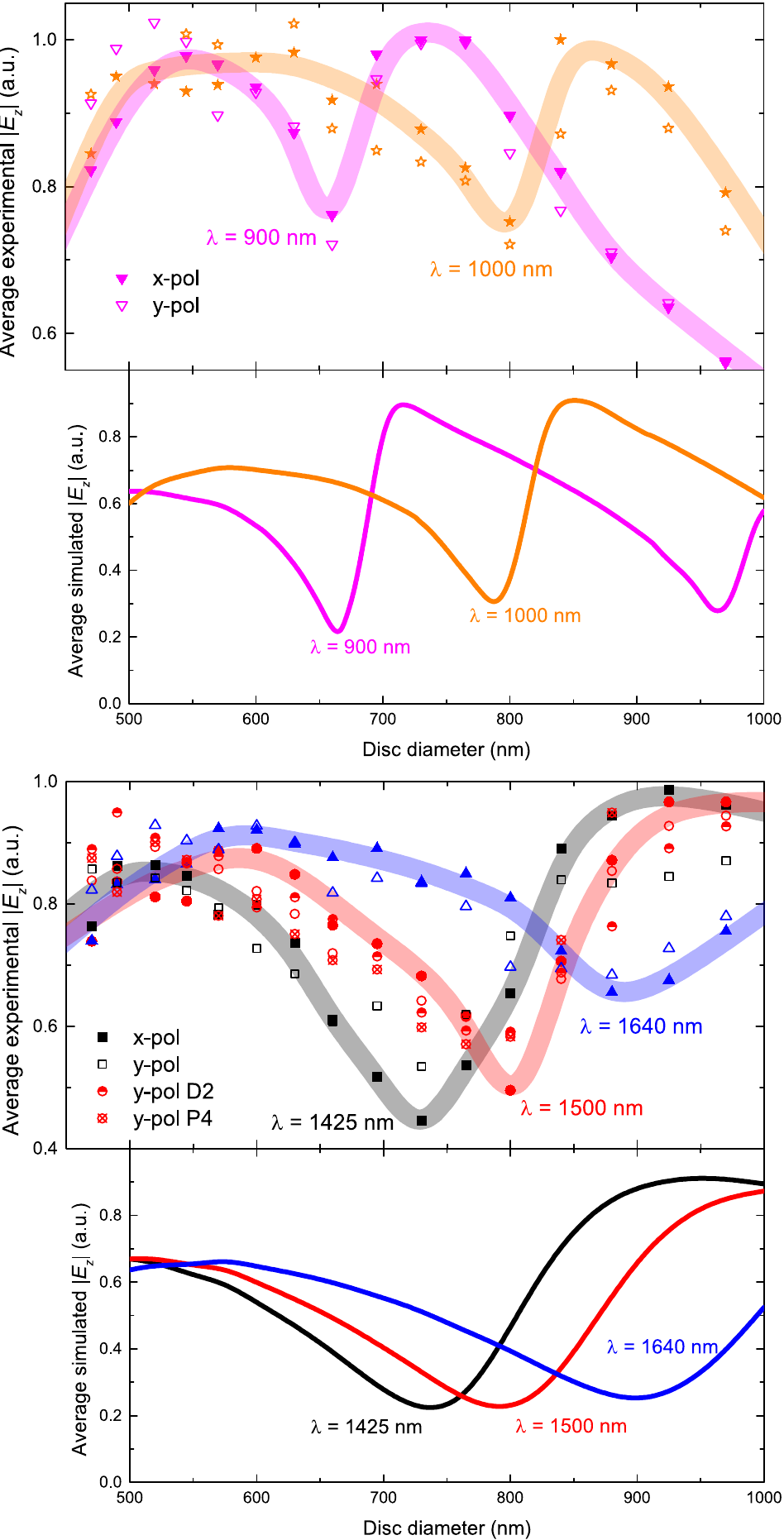}
  \caption{Comparison of measured and simulated average $\left|E_z\right|$. Simulations were done at altitude of 50 nm above disc surface. Agreement between experimental results for $x$- (solid) and $y$-polarized incident beam (hollow) proves the validity of the applied decomposition method. Agreement with results for structures with increased center-to-center disc separation of 4 $\mu$m (labeled P4) shows that the coupling effect is negligible. Finally, the agreement with results for a different duplicated structure (labeled D2) verifies the robustness of proposed anapole identification technique.}
	\label{S6}
\end{figure}

\end{document}